\documentclass[journal]{IEEEtran}
\usepackage{amsmath,amsfonts}

\usepackage[ruled,linesnumbered]{algorithm2e}

\usepackage{array}
\usepackage{textcomp}
\usepackage{stfloats}
\usepackage{url}
\usepackage{verbatim}
\usepackage{graphicx}
\usepackage{subfigure}
\usepackage{cite}
\usepackage{mathrsfs}
\usepackage{xcolor}
\usepackage{amssymb}
\usepackage{amsthm}
\usepackage{multirow}
\usepackage{threeparttable}
\usepackage{booktabs}
\usepackage{epstopdf}
\usepackage{stmaryrd}
\usepackage{bm}
\usepackage{mathbbol}

\newtheorem{theorem}{Theorem}
\newtheoremstyle{noparens}%
  {}{}%
  {\itshape}{}%
  {\bfseries}{.}%
  { }%
  {\thmname{#1}\thmnumber{ #2}\mdseries\thmnote{ #3}}
\theoremstyle{noparens}

\newtheorem{corollary}[theorem]{Corollary}

\newtheorem{definition}{Definition}

\begin{document}

\title{Linear Complexity Computation of Code Distance and Minimum Size of Trapping Sets for LDPC Codes with Bounded Treewidth}

\author{\IEEEauthorblockN{Qingqing Peng, Ke Liu, Guiying Yan, Guanghui Wang}

\thanks{This work is partially supported by the National Key L\&D Program of China, (2023YFA1009602)

Qingqing Peng is with the School of Mathematics, Shandong University, Jinan, 250100 China (e-mail:pqing@mail.sdu.edu.cn).

Ke Liu is with Huawei Technologies Co. Ltd., Hangzhou, 310051 China (e-mail: liuke79@huawei.com).

Guiying Yan is with the Academy of Mathematics and Systems Science, CAS, University of Chinese Academy of Sciences, Beijing, 100190 China (e-mail: yangy@amss.ac.cn).

Guanghui Wang is with the School of Mathematics, Shandong University, Jinan, 250100 China, and also with the State Key Laboratory of Cryptography and Digital Economy Security, Shandong University, Jinan, 250000 China (e-mail: ghwang@sdu.edu.cn).
}
}



\maketitle
\begin{abstract}
It is well known that, given \(b\ge 0\), finding an $(a,b)$-trapping set with the minimum \(a\) in a binary linear code is NP-hard. In this paper, we demonstrate that this problem can be solved with linear complexity with respect to the code length for codes with bounded treewidth. Furthermore, suppose a tree decomposition corresponding to the treewidth of the binary linear code is known. In that case, we also provide a specific algorithm to compute the minimum \(a\) and the number of the corresponding \((a, b)\)-trapping sets for a given \(b\) with linear complexity. Simulation experiments are presented to verify the correctness of the proposed algorithm.
\end{abstract}


\section{Introduction}
 
 Low-density parity-check (LDPC) codes \cite{gallager1962low} have been extensively studied by coding theorists and practitioners over the past decade because of their near-capacity performance when decoded using low-complexity iterative decoders. However, the error floor phenomenon at high signal-to-noise ratios (SNRs) causes the bit error rate (BER) curve to flatten, limiting further performance improvement. To address this issue, researchers have focused on identifying the underlying causes of the error floor. Extensive experimental studies \cite{richardson2003error,butler2014error,di2002finite,mackay2003weaknesses,dolecek2009analysis,vontobel2007graph} have shown that structures known as \((a,b)\)-trapping sets are the primary contributors to the error floor, where \(a\) represents the number of variable nodes in the structure, and \(b\) denotes the number of check nodes with an odd degree. In particular, trapping sets with small values of \(a\) and \(b\) are the most significant contributors to the error floor.

Identifying the smallest trapping set in an LDPC code is a crucial problem, which can be formulated as follows:

\textbf{Problem}: MINIMUM TRAPPING SET

\textbf{Instance}: A Tanner graph $G$, and an intege $b\ge 0$.

\textbf{Question}: What is the minimum value of \( a \) for which an \((a, b)\)-trapping set exists in graph \( G \)? And find an $(a,b)$-trapping set with minimum $a$.

\noindent In particular, when \( b = 0 \), finding the minimum size of an \((a,0)\)-trapping set is equivalent to computing the minimum distance of the LDPC code. Let \( H \) be the parity-check matrix corresponding to graph \( G \). If there exists an \((a,0)\)-trapping set \( S \) in \( G \), let \( \boldsymbol{x} \) be the vector where the bits corresponding to the variable nodes in \( S \) are set to 1, while all other bits are set to 0. It is straightforward to verify that \( H\boldsymbol{x} = \boldsymbol{0} \), and the weight of \( \boldsymbol{x} \) is \( a \). On the other hand, if \( \boldsymbol{x} \) is a weight-\( a \) vector satisfying \( H\boldsymbol{x} = \boldsymbol{0} \), let \( S \) be the set of variable nodes corresponding to the nonzero entries of \( \boldsymbol{x} \). In the induced subgraph of \( S \), each check node is connected to an even number of nodes in \( S \). Therefore, \( S \) is an \((a,0)\)-trapping set.

The study of the MINIMUM TRAPPING SET and its special cases has a long history. As early as 1971, Dominic Welsh \cite{welsh1971combinatorial} called for the development of efficient algorithms for computing the minimum distance of a code. For nearly two decades, this problem remained unsolved and was widely recognized as an open problem\cite{van1978inherent,garey1979computers,vardy1997intractability,johnson1986np}. It was not until 1997 that Alexander Vardy formally proved \cite{vardy1997intractability} that computing the minimum distance of a binary linear code is NP-hard. The study of more generalized trapping sets originated from Richardson's semi-analytical analysis in \cite{richardson2003error}, where he demonstrated that trapping sets are the primary cause of the error floor and proposed a method for rapidly estimating the error floor based on trapping sets. Since then, researchers have recognized the importance of identifying trapping sets in LDPC codes and have made considerable efforts to develop efficient methods for detecting dominant trapping sets \cite{kyung2010exhaustive,abu2010trapping,wang2007exhaustion,wang2009finding}. Unfortunately, as proven by McGregor et al. \cite{krishnan2006hardness,krishnan2007computing,mcgregor2010hardness}, not only is the problem of finding a minimum-size trapping set NP-hard, but even approximating its size remains NP-hard, regardless of the sparsity of the underlying graph. This implies that, for binary linear codes, no polynomial-time algorithms exist to solve the MINIMUM TRAPPING SET problem in the worst case.

Although the MINIMUM TRAPPING SET problem has been proven to be NP-hard, it continues to attract significant attention. Researchers have been persistently working on designing efficient algorithms to reduce the coefficient of exponential complexity, enabling the identification of the minimum size of trapping sets for codes of practical length (1000 to 10000) \cite{karimi2011efficient,karimi2012efficient,hashemi2016new,hashemi2018characterization}. Several reasons underscore the importance of this problem. Firstly, based on trapping sets, the error floor of LDPC codes can be quickly estimated \cite{richardson2003error,butler2014error}, which is particularly valuable for LDPC codes with error floors too low to be accurately observed through conventional computer simulations. Secondly, by modifying the decoder to account for the influence of specific trapping set structures, the decoder’s ability to mitigate error-prone patterns can be enhanced, thereby improving the overall error-correcting performance of the code(see, e.g., \cite{cavus2005performance} and \cite{han2008ldpc}). Furthermore, a better understanding of trapping sets can guide the design of LDPC codes by eliminating small trapping sets to achieve a lower error floor (see, e.g., \cite{karimi2019construction}).

The treewidth of a graph is a fundamental concept in graph theory. It indicates how far a graph is from being a tree or forest: the closer the graph is to a forest, the smaller its treewidth. The concept was first introduced by Bertel$\acute{\text{e}}$ et al. \cite{bertele1972nonserial} under the name dimension. It was later rediscovered by Halin \cite{halin1976s} in 1976 and by Robertson et al. \cite{robertson1984graph} in 1984. Now, treewidth has garnered significant attention due to its pivotal role in addressing classical NP-complete problems (see, for example, \cite{grohe2009tree,halin1976s,kloks1994treewidth,korach1993tree}). Notably, many NP-complete problems can be efficiently solved in polynomial time when restricted to graphs with bounded treewidth \cite{bodlaender1993linear}. One such example is the $w$-Dominating-Set problem, which can be formulated as follows.

\textbf{Problem}: $w$-Dominating-Set

\textbf{Instance}: Graph $G=(\mathcal{V},\mathcal{E})$, $k,  r\in \mathbb{N}$.

\textbf{Parameter}: treewidth $k$

\textbf{Question}: Let a $w$-dominating set be a vertex subset $S$ such that for all $v \in \mathcal{V}$, either $v \in S$ or it has at least $w$ neighbors in $S$. Decide whether $G$ has a $w$-dominating set of cardinality at most $r$. 

\noindent The MINIMUM TRAPPING SET problem, as an NP-hard problem, is also expected to be solvable in polynomial time with bounded treewidth.

Our primary contribution is proving that the MINIMUM TRAPPING SET problem can be solved with linear complexity in the code length for LDPC codes with bounded treewidth \(k\). Furthermore, if a tree decomposition of width \(k\) for the LDPC code is known, we also provide a specific algorithm to compute the size and the number of the smallest trapping sets for a given \(b\ge 0\) with linear complexity.

The paper is organized as follows. Section \ref{section_preliminaries} introduces the fundamental concepts of trapping sets in LDPC codes and treewidth. Section \ref{section_a0} demonstrates that the minimum distance can be computed with linear complexity when the treewidth is bounded. Section \ref{section_ab} shows that the minimum size of an \((a, b)\)-trapping set and the corresponding number of such sets for a given \(b\) can also be computed with linear complexity under bounded treewidth. Section \ref{section_algorithm} provides verification using spatially coupled LDPC codes as an example. Concluding remarks are provided in Section \ref{section_conclusion}.

\section{PRELIMINARIES}
\label{section_preliminaries}

In this section, we define several important concepts, such as trapping sets and treewidth, as well as some set operations that are essential for Sections \ref{section_a0} and \ref{section_ab}.

\subsection{Trapping Set}
Let \( G = (\mathcal{L} \cup \mathcal{R}, \mathcal{E}) \) be a bipartite graph, or Tanner graph, corresponding to the LDPC code, where \( \mathcal{L} \) is the set of variable nodes, \( \mathcal{R} \) is the set of check nodes, and \( \mathcal{E} \) represents the edges between \( \mathcal{L} \) and \( \mathcal{R} \). For a node \( u \in \mathcal{L} \cup \mathcal{R} \), its degree is the number of neighbors of \( u \). For a subset \( S \subset \mathcal{L} \), let \(\Gamma_G(S)\) denote the set of neighbors of \(S\) in \(G\). The induced subgraph of $S \cup \Gamma_G(S)$ in the Tanner graph, denoted \( G[S \cup \Gamma_G(S)] \), has \( S \cup \Gamma_G(S) \) as the set of vertices, and the set of edges is given by \( \{(v, c) | v \in S, c \in \Gamma_G(S), (v, c) \in \mathcal{E}\} \). Let \(\Gamma_G^o(S)\) denote the set of check nodes in \(\Gamma_G(S)\) with odd degree in \(G([S \cup \Gamma_G(S)])\). For notational simplicity, throughout this paper we also use \(G[S]\) to denote \(G[S \cup \Gamma_G(S)]\).

\begin{definition} 
For a Tanner graph \( G = (\mathcal{L} \cup \mathcal{R}, \mathcal{E}) \) and a non-empty subset \( S \subset \mathcal{L} \), we define \( S \) as an \( (a, b) \) trapping set in $G$ if \( |S| = a \) and \( |\Gamma_G^o(S)| = b \). The integer \( a \) is referred to as the size of the trapping set \( S \).
\end{definition}

\subsection{Treewidth}
Tree decomposition and treewidth are important concepts in graph theory \cite{alber2002fixed}. LDPC codes with bounded treewidth are a class of codes that this paper focuses on. 

Consider a bipartite graph \( G = (\mathcal{L} \cup \mathcal{R}, E) \). Let \( T \) be a tree, and \( (B_t)_{t \in T} \) be a collection of node subsets \( B_t \subset \mathcal{L} \cup \mathcal{R} \) indexed by the nodes of \( T \). Then, the tree decomposition can be defined as follows:
\begin{definition}
\label{tree_decomposition}
The tree decomposition of \( G = (\mathcal{L} \cup \mathcal{R}, \mathcal{E}) \) is a pair \( (T, (B_t)_{t \in T}) \) such that:
\begin{enumerate}
\item \( \mathcal{L} \cup \mathcal{R} = \cup_{t \in T} B_t \);
\item for every edge \( (v, c) \in \mathcal{E} \), there exists a \( t \in T \) such that \( v,c \in B_t \);
\item for every node \( u \in \mathcal{L} \cup \mathcal{R}\), the set \( B^{-1}(u) = \{ t \in T | u \in B_t \} \) is non-empty and connected in \( T \).
\end{enumerate}
\end{definition}
The width of a tree decomposition \( (T, (B_t)_{t \in T}) \) is defined as 
\[
   \max_{t\in T} |B_t|-1. 
\]
The treewidth of a graph \( G \), denoted as \(tw(G) \), is the minimum width among all tree decompositions of \( G \). The pathwidth of a graph \( G \), denoted as \( pw(G) \), is the minimum width among all tree decompositions of \( G \) in which \( T \) is a path.

\begin{figure}[!t]
\centering
\subfigure[A bipartite graph \( G \) and a tree decomposition of \( G \) with width 3.]{
\includegraphics[width=2in]{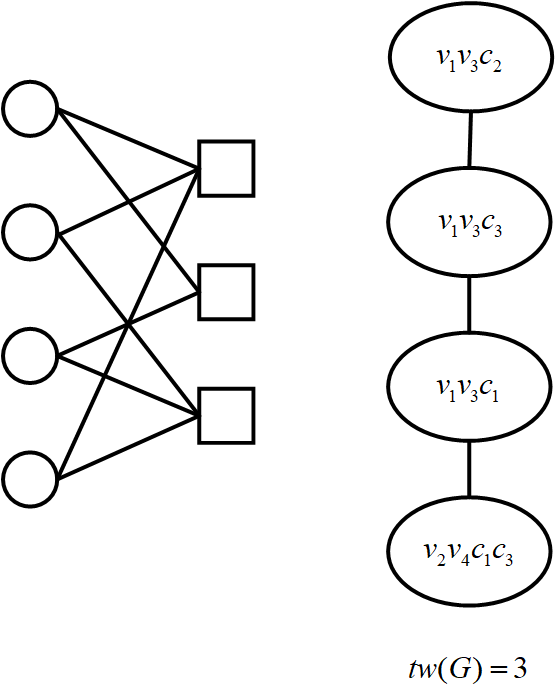} 
\label{a}
}
\subfigure[A tree decomposition of \( G \) with width 2.]{
\includegraphics[width=2in]{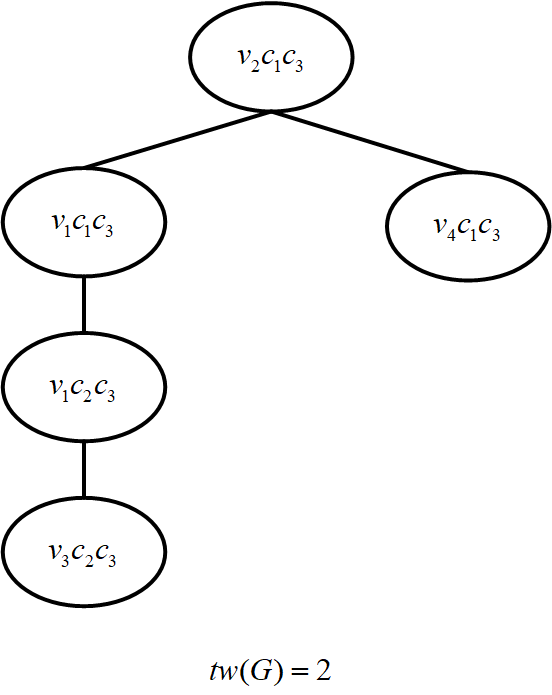} 
\label{b}
}
\DeclareGraphicsExtensions.
\caption{We depict a bipartite graph \( G \) with \( tw(G) = 2 \), along with its two tree decompositions. Different tree decompositions may have different widths, but we are primarily concerned with the tree decomposition of minimum width.}
\label{define_tw}
\end{figure}

As defined in Definition \ref{tree_decomposition}, for each graph \( G = (\mathcal{L} \cup \mathcal{R}, \mathcal{E}) \), there exists a tree decomposition \( (T, (B_t)_{t \in T}) \) where \( T \) contains only a single node \( t \) such that \( B_t = \mathcal{L} \cup \mathcal{R} \). This implies that the treewidth of \( G \) is at most \( |\mathcal{L} \cup \mathcal{R}| - 1 \). A rooted tree decomposition is
a tree decomposition with a distinguished root node. For example, Fig.\ref{define_tw} illustrates two rooted tree decompositions on a bipartite graph.

To better analyze algorithms based on tree decomposition, the concept of rooted nice tree decomposition \cite{alber2002fixed} is essential.
\begin{definition}
\label{def_3}
We say that the rooted tree decomposition \( (T, (B_t)_{t \in T}) \) is nice if it satisfies the following properties:
\begin{enumerate}
\item every node in \( T \) has at most two child nodes;
\item if a node \( t \in T \) has two child nodes \( t' \) and \( t'' \), then \( B_t = B_{t'} = B_{t''} \). node \( t \) is called a \textbf{join node};
\item if a node \( t \in T \) has a single child node \( t' \), then one of the following two conditions must hold:
\begin{itemize}
\item There exists a vertex \( u \in \mathcal{L} \cup \mathcal{R} \) such that \( B_t = B_{t'} \cup \{u\} \). In this case, node \( t \) is called an \textbf{introduced node};
\item There exists a vertex \( u \in \mathcal{L} \cup \mathcal{R} \) such that \( B_t = B_{t'} \setminus \{u\} \). In this case, node \( t \) is called a \textbf{forget node}.
\end{itemize}
\end{enumerate}
\end{definition}

\begin{figure*}[htbp]
\centering
\includegraphics[width=1.0\textwidth]{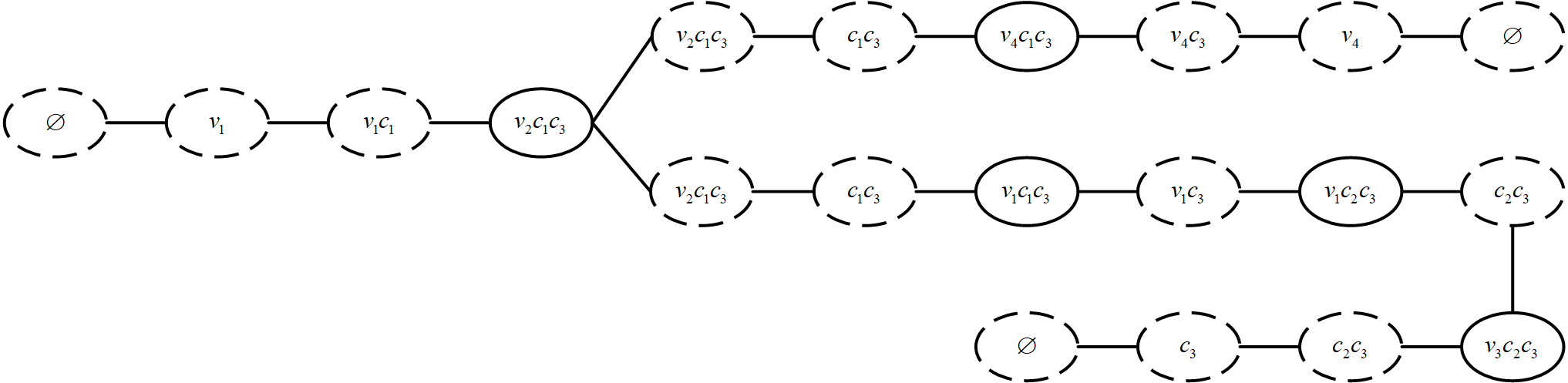}
\caption{The rooted nice tree decomposition corresponding to the bipartite graph in Fig.\ref{a}. This nice tree decomposition can be obtained by adding some ``auxiliary nodes" (dashed lines) to the tree decomposition in Fig.\ref{b}.}
\label{nice_tree_decomposition}
\end{figure*}

It is not hard to transform a given tree decomposition into a rooted nice tree decomposition(see \cite[Lemma 13.1.3]{kloks1994treewidth}). As shown in Fig.\ref{nice_tree_decomposition}, it is a rooted nice tree decomposition transformed from Fig.\ref{b}, with the same width preserved. In the following discussion of tree decompositions in this paper, we assume that the tree decomposition is in rooted nice form.

\subsection{Set Operations}
\label{set}
Let \( A \) and \( B \) be two sets. The fundamental set operations, including power set, intersection, union, difference, and symmetric difference, are defined as follows:  
\begin{itemize}
\item Intersection (\(\cap\)): The intersection of \( A \) and \( B \), denoted as \( A \cap B \), is the set of elements that belong to both \( A \) and \( B \), i.e.,  
  \[
  A \cap B = \{ x \mid x \in A \text{ and } x \in B \}.
  \]

\item Union (\(\cup\)): The union of \( A \) and \( B \), denoted as \( A \cup B \), is the set of elements that belong to at least one of \( A \) or \( B \), i.e.,  
  \[
  A \cup B = \{ x \mid x \in A \text{ or } x \in B \}.
  \] 
\item Difference (\(\setminus\)): The difference of \( A \) and \( B \), denoted as \( A \setminus B \), is the set of elements that belong to \( A \) but not to \( B \), i.e.,  
  \[
  A \setminus B = \{ x \mid x \in A \text{ and } x \notin B \}.
  \]  
\item Symmetric Difference (\(\oplus\)): The symmetric difference of \( A \) and \( B \), denoted as \( A \oplus B \), is the set of elements that belong to either \( A \) or \( B \), but not both, i.e.,  
  \[
  \begin{split}
   A \oplus B &= (A \setminus B) \cup (B \setminus A) \\
   &= \{ x \mid x \in A \text{ or } x \in B, \text{ but } x \notin A \cap B \}.  
  \end{split} 
  \] 
  
\end{itemize}

Let \( U \) and \( W \) be two sets. These operations satisfy several fundamental properties:  

\noindent Commutativity:  
  \[
  A \cup B = B \cup A, \quad A \cap B = B \cap A, \quad A \oplus B = B \oplus A.
  \]  
\noindent associativity:  
  \[
  (A \cup B) \cup U = A \cup (B \cup U), \quad (A \cap B) \cap U = A \cap (B \cap U).
  \]  
\noindent Distributive Laws:  
  \[
  A \cap (B \cup U) = (A \cap B) \cup (A \cap U),
  \]  
  \[
  A \cup (B \cap U) = (A \cup B) \cap (A \cup U).
  \]
Other operations involving \(\oplus\) and \(\setminus\) are more complex. Here, we list several operational rules that will be used in subsequent sections:
\begin{itemize}
    \item $(A \oplus B) \cap U =(A \cap U) \oplus (B \cap U)$,
    \item $(A \setminus B) \cap U = (A\cap U) \setminus (B \cap U)$,
    \item $(A \setminus B ) \cap (U \setminus W) = (A \cap U) \setminus (B \cap W)$,
    \item$A \setminus(B\setminus C) = (A \setminus B) \cup (A \cap C)$
    \item $(A \oplus B)\setminus C =(A \setminus(B \cup C))\cup (B \setminus(A \cup C))$ 
    \item $A \setminus(B \cup C) = (A\setminus B)\setminus C$
\end{itemize}


\begin{figure}[htbp]
\centering
\includegraphics[width=0.27\textwidth]{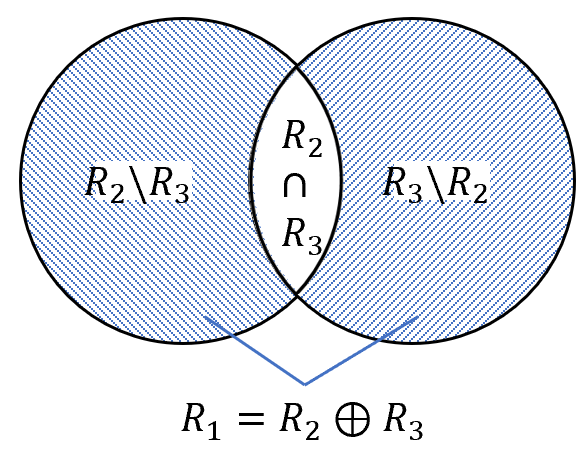}
\caption{A Venn diagram illustrating the symmetric difference and intersection of sets \( R_2 \) and \( R_3 \).}
\label{venn}
\end{figure}

\section{Computer the Size and Number of the Smallest Trapping Set with \( b = 0 \) in Codes with Bounded Treewidth}
\label{section_a0}
Consider a bipartite graph \(G = (\mathcal{L} \cup \mathcal{R}, \mathcal{E})\) corresponding to the LDPC code \(\mathcal{C}\), with treewidth \(k\). Next, for a given \( b \), we will prove that finding an \((a, b)\)-trapping set with the minimum \( a \) in \( G \) can be solved in linear complexity with respect to the code length. In this section, we focus on the case where \( b = 0 \), while the generalization to \( b > 0 \) will be presented in Section \ref{section_ab}. The motivation for first considering the case \( b = 0 \) is twofold. On the one hand, the \((a,0)\)-trapping set is the influential structure affecting the error floor of LDPC codes, and finding the smallest trapping set with $b=0$ in \( G \) has received significant attention. On the other hand, the case \( b > 0 \) is more complex. To enhance clarity and facilitate understanding, we start with the simpler case of \( b = 0 \). Furthermore, we also show that the number of smallest trapping sets with $b=0$ contained in \( G \) can be computed in linear complexity with respect to the code length.
\subsection{Several Important Properties of Tree Decomposition}

To better design algorithms for finding the smallest trapping set, we introduce several important properties to deepen our understanding of trapping set.

\textbf{Property 1:} Let \( S, S_1, S_2 \) be three trapping sets in \( G\), where \( S = S_1 \cup S_2 \) and \( S_1 \cap S_2 = \emptyset \). Then, 
\[
\Gamma^o_G(S) = \Gamma^o_G(S_1) \oplus \Gamma^o_G(S_2).
\]
and
\[
\Gamma^o_G(S_1) = \Gamma^o_G(S) \oplus \Gamma^o_G(S_2).
\]

\textbf{Property 2:} Let \( S \) be a trapping set in \( G\), and let $\mathcal{L}'$ be a check node sets in $G$. 
\[
\Gamma^o_G(S) = \Gamma^o_{G\setminus \mathcal{L}'}(S) \oplus \Gamma^o_{G[\mathcal{L}'\cup S]}(S).
\]

\subsection{Compute The Size and Number of The Smallest Trapping Set with $b=0$}
Before presenting the algorithms for finding the smallest trapping set, we first introduce the following lemma, which plays a critical role in the proof.

Next, we present an algorithm to compute the size and the number of the smallest trapping set with \( b = 0 \) in $G$, along with its computational complexity. Additionally, we provide a detailed proof to verify the correctness of the algorithm.

\begin{theorem}
\label{main_a0}
Let \( G \) be a Tanner graph with treewidth \( tw(G)=k \), corresponding to an LDPC code of length \( n \). Then, for given $b=0$, the size and the number of the smallest $(a,0)$ trapping sets in \( G \) can be determined in time \( O(n 4^{k}) \). 
\end{theorem}

\begin{proof}
A nice tree decomposition of \( G \) is given by \( (T, (B_t)_{t \in T}) \) with width \( k \). Without loss of generality, we may assume that each bag \( B_t \) corresponding to a leaf node or the root node in \( T \) is an empty set. Based on this decomposition, we develop a dynamic programming algorithm that solves the MINIMUM TRAPPING SET problem.


The algorithm processes the nodes of $T$ in a bottom-up order $t_1,t_2,\ldots,t_r$, such that every child of a node $t_i$ appears earlier in the order. At each node $t_i$, the algorithm maintains information about trapping sets contained in the subgraph induced by the union of the bags processed so far, namely $G_i:=G[\cup_{l=1:i}B_{t_l}]$. Let \(\mathcal{L}_i\) denote the set of variable nodes in \(G_i\), and let $\mathcal{L}^B_i$ and $\mathcal{R}^B_i$ denote the the sets of variable nodes and check nodes in $B_{t_i}$, respectively.

\begin{definition} Let \(S\) be a trapping set in \(G\), \(L\) a set of variable nodes, \(I\) a set of check nodes. We say that \(S\) is an \((L, I, t_i)\)-type trapping set if the following conditions hold:
\begin{enumerate}
    \item $S \subset\mathcal{L}_i$,
    \item $S \cap B_{t_i} = L$,
    \item $\Gamma^o_{G_i}(S)=I$,
\end{enumerate}
\end{definition}

Formally, for each node \( t_i \in T \), we define two functions as follows,
\[
f_{t_i} : (L , I) \mapsto f_{t_i}(L, I) \in \mathbb{Z},
\]
and
\[
g_{t_i} : (L, I) \mapsto g_{t_i}(L, I) \in \mathbb{Z}.
\]
where $L \subset \mathcal{L}^B_i$ is a set of variable nodes in $B_{t_i}$, $I\subset \mathcal{R}^B_i$ is a set of check nodes in $B_{t_i}$. The value $f_{t_i}(L, I)$ denotes the minimum size of a $(L,I,t_i)$-type trapping set, while $g_{t_i}(L, I)$ counts the number of such minimum-size trapping sets.


The value of $f_{t_i}$ and $g_{t_i}$ are updated according to the type of node $t_i$ in the nice tree decomposition, using the values computed at its children. After processing root node $t_r$, \(f_{t_r}(\emptyset,\emptyset)\) gives the minimum size of an \((a,b)\) trapping set contained in $G$, while \(g_{t_r}(\emptyset,\emptyset)\) gives the number of minimum \((a,b)\) trapping set contained in $G$. The detailed update rules and complexity analysis are provided in the sequel.

\noindent  \textbf{Leaf node}:  Suppose $t_i$ is a leaf node in $T$. Then,
\[
    f_{t_i}(L,I) = +\infty, 
\]
and
\[
    g_{t_i}(L,I) = 0,
\]
for $L=\emptyset$, $I =\emptyset$. It takes \( O(1) \) time to compute all \( f_{t_i}(L, I) \) and \( g_{t_i}(L, I) \) for each leaf node \( t_i \).

\noindent \textbf{introduced variable node}: Suppose \( t_i \) is an introduced node with a unique child \( t_j \), and that the corresponding bags satisfy \( B_{t_i} = B_{t_j} \cup \{v\} \), where \( v \) is a variable node. We can compute \( f_{t_i} \) and $g_{t_i}$ using the equations (\ref{a0_introduce_variable_f}) and (\ref{a0_introduce_variable_g}) for all variable node set $L \subset {\mathcal{L}^B_{i}}$, and check nodes set $I \subset {\mathcal{R}^B_{i}}$.
\begin{figure*}
    \begin{equation}
\label{a0_introduce_variable_f}
   f_{t_i}(L,I) = \begin{cases}
        f_{t_j}(L,I),  &\text{if } v \notin L; \\ 
        1,  &\text{if } L = \{v\} , \text{ and } I = \Gamma_{G_i}^o(v); \\
        f_{t_j}(L\backslash  \{v\},I \oplus \Gamma_{G_i}^o(v)) + 1,  &\text{otherwise.}
   \end{cases}
\end{equation}
\end{figure*}
\begin{figure*}
\begin{equation}
\label{a0_introduce_variable_g}
   g_{t_i}(L,I) = \begin{cases}
        g_{t_j}(L,I),  &\text{if } v \notin L; \\ 
        1,  &\text{if } L = \{v\} , \text{ and } I = \Gamma_{G_i}^o(v); \\
        g_{t_j}(L\backslash  \{v\},I \oplus \Gamma_{G_i}^o(v)),  &\text{otherwise.}
   \end{cases}
\end{equation}
\end{figure*}

Next, we prove the correctness of (\ref{a0_introduce_variable_f}) and (\ref{a0_introduce_variable_g}). Let \( S \) be a trapping set in \( G \). For each pair \( (L , I) \), the following conclusion holds.
\begin{itemize}
    \item If $v \notin L$, $S$ is a \( (L, I,t_i) \)-type trapping set if and only if $S$ is also a \( (L, I,t_j) \)-type trapping set, which leads to
$f_{t_i}(L, I) = f_{t_j}(L, I) $ and $g_{t_i}(L, I) = g_{t_j}(L, I)$. This is because, when \( v \notin S \), the following equation holds:
\begin{equation}
\label{a0_ivn_1}
\begin{split}
    S \subset\mathcal{L}_j\Leftrightarrow S \subset \mathcal{L}_i
\end{split}
\end{equation}
\begin{equation}
\label{a0_ivn_2}
\begin{split}
    S \cap B_{t_j} &=  S \cap (B_{t_i}\setminus\{v\})\\
    &=(S\cap B_{t_i})\setminus (S \cap \{v\})\\
    &=S\cap B_{t_i}
\end{split}
\end{equation}
and
\begin{equation}   
\label{a0_ivn_3}
\Gamma_{G_{j}}^o(S) = \Gamma_{G_i}^o(S).
 \end{equation}

When \( S \) is a \( (L, I, t_i) \)-type trapping set, we have \( S \cap B_{t_i} = L \), and \( v \notin L \), thus \( v \notin S \). When \( S \) is a \( (L, I, t_j) \)-type trapping set, we have \( S \subset\mathcal{L}_j\), thus \( v \notin S \).

 \item If \( L = \{v\} \) and \( I = \Gamma^o_{G_i}(v) \), then \( \{v\} \) is a \( (L, I,t_i) \)-type trapping set. In this case, we have \( f_{t_i}(L, I) = 1 \) and \( g_{t_i}(L, I) = 1 \).

\item For other cases, \( S \) has the form \( S = \{v\} \cup S' \), where $v \notin S'$. We claim $S$ is a \( (L, I,t_i) \)-type trapping set if and only if $S'$ is a \( (L\setminus \{v\}, I \oplus \Gamma^o_{G_i}(v),t_j) \)-type trapping set, which leads to
$f_{t_i}(L,I) = f_{t_j}(L\setminus \{v\} , I \oplus \Gamma^o_{G_i}(v))+1.$ and $g_{t_i}(L,I) = g_{t_j}(L\setminus \{v\} , I \oplus \Gamma^o_{G_i}(v))$. According to $B_{t_i} = B_{t_j} \cup \{v\}$, we have:
\begin{equation}
\label{a0_ivn_4}
\begin{split}
    S' \subset\mathcal{L}_j\Leftrightarrow S \subset \mathcal{L}_i
\end{split}
\end{equation}

\begin{equation}
\label{a0_ivn_5}
\begin{split}
    S'\cap B_{t_j} &= (S\setminus\{v\}) \cap (B_{t_i}\setminus \{v\})\\
    &= (S\cap B_{t_i}) \setminus (\{v\} \cap \{v\}) \\
    &=(S\cap B_{t_i}) \setminus \{v\}
\end{split}
 \end{equation}
and
\begin{equation}
\label{a0_ivn_6}
\begin{split}
    \Gamma^o_{G_j}(S') &= \Gamma^o_{G_i}(S\setminus\{v\}) \\
    &\overset{prop.1}{=} \Gamma^o_{G_i}(S) \oplus \Gamma^o_{G_i}(v)
\end{split}
 \end{equation}

Clearly, based on (\ref{a0_ivn_4}) $\sim$ (\ref{a0_ivn_6}), the sufficiency of the claim is established. Conversely, to establish necessity, suppose \( S' \) is a \( (L \setminus {v}, I \oplus \Gamma^o_{G_i}(v), t_j) \)-type trapping set, then \( S \cap B_{t_i} = L \) or \( S \cap B_{t_i} = L\setminus\{v\} \). Since \( v \in S\), it follows that \( S \cap B_{t_i} = L \). Therefore, \( S \) is a \( (L, I, t_i) \)-type trapping set.
\end{itemize}

It takes \( O(2^k) \) time for each introduced variable node to compute \( f_{t_i}(L, I) \) and \( g_{t_i}(L, I) \) for all \( (L, I) \).

\noindent \textbf{Forget variable node}: Suppose \( t_i \) is a forget node with a unique child \( t_j \), and that the corresponding bags satisfy \( B_{t_i} = B_{t_j} \setminus \{v\} \), where \( v \) is a variable node. We can compute \( f_{t_i} \) and \( g_{t_i} \) using the following equations for all $L \subset {\mathcal{L}^B_{i}}$, $I \subset {\mathcal{R}^B_{i}}$:
\begin{equation}
\label{a0_forget_variable_f}
    f_{t_i}(L,I) = \min \{f_{t_j}(L,I),f_{t_j}(L\cup \{v\},I)\}.
\end{equation}
and
\begin{equation}
\label{a0_forget_variable_g}
    g_{t_i}(L,I) =  \sum_{\substack{P\in\{(L,I),(L\cup \{v\},I)\}\\f_{t_i}(L,I)=f_{t_j}(P) }} g_{t_j}(P).
\end{equation}

Let \( S \) be a trapping set in \( G \). For each pair \( (L \subset \mathcal{L}^B_{i}, I \subset \mathcal{R}^B_{i}) \), we claim that \( S \) is a \( (L, I,t_i) \)-type trapping set if and only if $S$ is either a \( (L, I,t_j) \)-type trapping set or a \( (L\cup \{v\}, I,t_j) \)-type trapping set, which leads to (\ref{a0_forget_variable_f}) and (\ref{a0_forget_variable_g}). According to $B_{t_i} = B_{t_j} \setminus \{v\}$, we have:
\begin{equation}
\label{a0_fvn_1}
  \begin{split}
    S \subset\mathcal{L}_j\Leftrightarrow S \subset \mathcal{L}_i
\end{split}  
\end{equation}
\begin{equation}
\label{a0_fvn_2}
  \begin{split}
    S \cap B_{t_j} &= S \cap (B_{t_i}\cup \{v\})\\
    &=(S\cap B_{t_i}) \cup (S \cap \{v\}),
\end{split}  
\end{equation}
and
\begin{equation}
\label{a0_fvn_3}
 \begin{split}
     \Gamma^o_{G_j}(S) =\Gamma^o_{G_i}(S).
\end{split}   
\end{equation}
Clearly, based on equations (\ref{a0_fvn_1}) $\sim$ (\ref{a0_fvn_3}), the sufficiency of the claim is established. Conversely, to establish necessity, suppose \( S \) is a \( (L, I, t_j) \)-type trapping set. Since \( v \notin L,v \in B_{t_j} \), and $S \cap B_{t_j} = L$, we have \( S \cap \{v\} = \emptyset \), which implies that \( S \cap B_{t_i} = L \). Therefore, \( S \) is a \( (L, I, t_i) \)-type trapping set. Similarly, suppose \( S \) is a \( (L \cup \{v\}, I, t_j) \)-type trapping set. According to \eqref{a0_fvn_2}, \( S \cap B_{t_i} = L \) or \( S \cap B_{t_i} = L \cup \{v\} \). Since \( v \notin B_{t_i} \), it follows that \( S \cap B_{t_i} = L \), and thus \( S \) is a \( (L, I, t_i) \)-type trapping set.

It takes \(  O(2^k) \) time for each forget variable node to compute \( f_{t_i}(L, I) \) and \( g_{t_i}(L, I) \) for all \( (L, I) \).

\noindent \textbf{introduced check node}: Suppose \( t_i \) is an introduced node with a unique child \( t_j \), and that the corresponding bags satisfy \( B_{t_i} = B_{t_j} \cup \{c\} \), where \( c \) is a check node. We can compute \( f_{t_i} \) and \( g_{t_i} \) using the following equations for all $L \subset {\mathcal{L}^B_{i}}$, $I \subset {\mathcal{R}^B_{i}}$:

\begin{equation}
\label{a0_introduce_check_f}
f_{t_i}(L,I) = \begin{cases}
f_{t_j}(L,I\setminus\{c\}), & \text{if } c \in I \text{, $E(c,L)$ is odd;}\\
f_{t_j}(L,I),& \text{if } c \notin I \text{, $E(c,L)$ is even;}\\
+\infty, & \text{otherwise},
\end{cases}
\end{equation}
\begin{equation}
\label{a0_introduce_check_g}
g_{t_i}(L,I) = \begin{cases}
g_{t_j}(L,I\setminus\{c\}), & \text{if } c \in I \text{, $E(c,L)$ is odd;}\\
g_{t_j}(L,I),& \text{if } c \notin I \text{, $E(c,L)$ is even;}\\
0, & \text{otherwise},
\end{cases}
\end{equation}
where \( E(c,L) \) denotes the number of edges between \( c \) and \( L \) in \( G_i\). 

Next, we prove the correctness of (\ref{a0_introduce_check_f}) and (\ref{a0_introduce_check_g}). Let \( S \) be a trapping set in \( G \). For each pair \( (L \subset \mathcal{L}^B_{i}, I \subset \mathcal{R}^B_{i}) \), we have:
\begin{equation}
\label{a0_icn_1}
\begin{split}
   S \subset\mathcal{L}_j\Leftrightarrow S \subset \mathcal{L}_i
\end{split}
\end{equation}
\begin{equation}
\label{a0_icn_2}
\begin{split}
    S \cap B_{t_j} =  S \cap B_{t_i}
\end{split}
\end{equation}
and
\begin{equation}   
\label{a0_icn_3}
\begin{split}
 \Gamma_{G_{j}}^o(S) & = \Gamma_{G_i \setminus \{c\}}^o(S)\\
&\overset{prop.2}{=} \Gamma_{G_i}^o(S) \oplus  \Gamma_{G[c\cup S]}^o(S) \\
&\overset{(a)}{=}\Gamma_{G_i}^o(S) \oplus  \Gamma_{G[c\cup S]}^o(S\cap B_{t_i})
\end{split}
 \end{equation}
 The validity of $(a)$ follows from the definition of treewidth: since \(c\) appears for the first time in \(B_{t_i}\), it can be adjacent only to variable nodes within \(B_{t_i}\), implying that \(E(c,S)=E(c, S\cap B_{t_i})\). Therefore, the following equation holds:
\begin{equation}
    \begin{split}
    &\Gamma_{G[c\cup S]}^o(S)\\
    &=\Gamma_{G[c\cup S]}^o(S\cap B_{t_i}) =\begin{cases}
    \{c\},&\text{if } E(c,S\cap B_{t_i}) \text{ is odd}\\
    \emptyset,&\text{if } E(c,S\cap B_{t_i}) \text{ is even}
\end{cases}
\end{split}
\label{a0_icn_4}
\end{equation}

 \begin{itemize}
\item If $E(c,L)$ is odd and $c \in I$, according to (\ref{a0_icn_1})–(\ref{a0_icn_4}), we have that $S$ is a \( (L, I,t_i) \)-type trapping set if and only if $S$ is a \( (L, I\setminus\{c\},t_j) \)-type trapping set. Consequently, $f_{t_i}(L, I) = f_{t_j}(L, I\setminus\{c\}) $ and $g_{t_i}(L, I) = g_{t_j}(L, I\setminus\{c\})$. 
 
\item If $E(c,L)$ is even and $c \notin I$, according to (\ref{a0_icn_1})–(\ref{a0_icn_4}), we have that $S$ is a \( (L, I,t_i) \)-type trapping set if and only if $S$ is a \( (L, I,t_j) \)-type trapping set. Consequently, $f_{t_i}(L, I) = f_{t_j}(L, I) $ and $g_{t_i}(L, I) = g_{t_j}(L, I)$.

\item If $E(c,L)$ is odd and $c \notin I$, we claim that there does not exist an $(L,I,t_i)$-type trapping set $S$ in $G_i$. Otherwise, since \(E(c,L=S\cap B_{t_i})\) is odd, we have $E(c,S)$ is odd and \(
c \in \Gamma^{o}_{G_i}(S)
\), which contradicts the assumption that \(c \notin I\). 

\item  If $E(c,L)$ is even and $c \in I$, we claim that there does not exist an $(L,I,t_i)$-type trapping set $S$ in $G_i$. Otherwise, Since \(E(c,L)\) is even, we have $E(c,S)$ is even and \(
c \notin \Gamma^{o}_{G_i}(S)
\), which contradicts the assumption that \(c \in I\). 
 \end{itemize}

It takes \( O(2^k) \) time for each introduced check node to compute \( f_{t_i}(L, I) \) and \( g_{t_i}(L, I) \) for all \( (L, I) \).

\noindent \textbf{Forget check node}: Suppose \( t_i \) is a forget node with a unique child \( t_j \), and that the corresponding bags satisfy \( B_{t_i} = B_{t_j} \setminus \{c\} \), where \( c \) is a check node. We can compute \( f_{t_i} \) and \( g_{t_i} \) using the following equations for all $L \subset {\mathcal{L}^B_{i}}$, $I \subset {\mathcal{R}^B_{i}}$:

\begin{equation}
\label{a0_forget_check_f}
f_{t_i}(L,I) = f_{t_j}(L,I),
\end{equation}
and 
\begin{equation}
\label{a0_forget_check_g}
g_{t_i}(L,I) = g_{t_j}(L,I).
\end{equation}

 Let \( S \) be a trapping set in \( G \). For each pair \( (L \subset \mathcal{L}^B_{i}, I \subset \mathcal{R}^B_{i}) \), we claim that $S$ is a \( (L, I,t_i) \)-type trapping set if and only if $S$ is a \( (L, I,t_j) \)-type trapping set, which leads to
$f_{t_i}(L, I) = f_{t_j}(L, I) $ and $g_{t_i}(L, I) = g_{t_j}(L, I)$. This claim follows from the fact that:
\begin{equation}
\label{a0_fcn_1}
\begin{split}
   S \subset\mathcal{L}_j\Leftrightarrow S \subset \mathcal{L}_i
\end{split}
\end{equation}
\begin{equation}
\label{a0_fcn_2}
\begin{split}
    S \cap B_{t_j} =  S \cap B_{t_i}
\end{split}
\end{equation}
and
\begin{equation}   
\label{a0_fcn_3}
\begin{split}
 \Gamma_{G_{j}}^o(S) &= \Gamma_{G_i }^o(S)
\end{split}
 \end{equation}

It takes \( O(2^k) \) time for each forget check node to compute \( f_{t_i}(L, I) \) and \( g_{t_i}(L, I) \) for all \( (L, I) \).

\noindent \textbf{Join node}: Suppose \( t_i \) is a join node, and \( t_{j_1} \) and \( t_{j_2} \) be its two child nodes with $B_{t_i} = B_{t_{j_1}} = B_{t_{j_2}}$. We define the following two functions: 

\begin{equation}
\label{a0_join_node_f1}
F_{t_i}(L,I) = \min_{\substack{I',I''}}\{f_{t_{j_1}}(L,I')+ f_{t_{j_2}}(L,I'') - |L|\}.
\end{equation}
\begin{equation}
\label{a0_join_node_g1}
\begin{split}
  &G_{t_i}(L,I) = \sum_{\substack{I',I''}} g_{t_{j_1}}(L,I') \times g_{t_{j_2}}(L,I'').
\end{split}
\end{equation}
where (\ref{a0_join_node_f1}) takes the minimum over all \( I', I''\) satisfying \( I = I' \oplus I'' \oplus \Gamma^o_{G[B_t]}(L)\), (\ref{a0_join_node_g1}) takes the sum over all \( I',I''\) that attain the minimum value in (\ref{a0_join_node_f1}). 
We can compute \( f_{t_i} \) and \( g_{t_i} \) using the following equations for all $L \subset {\mathcal{L}^B_{i}}$, $I \subset {\mathcal{R}^B_{i}}$:

\begin{equation}
\label{a0_join_node_f2}
f_{t_i}(L,I) =
\begin{cases}
     F_{t_{i}}(L,I), \text{\qquad\qquad\qquad\qquad\qquad\;\;\;\; if } L \neq \emptyset \\
     \min \{F_{t_i}(L,I),f_{t_{j_1}}(L,I), f_{t_{j_2}}(L,I) \}, \text{if } L = \emptyset
\end{cases}
\end{equation}
\begin{equation}
\label{a0_join_node_g2}
g_{t_i}(L,I) =
\begin{cases}
     G_{t_i}(L,I),& \text{if } L \neq \emptyset; \\
     \mathbb{1}_{f_{t_i}(L,I) = F_{t_i}(L,I)}G_{t_i}(L,I)
     \\+\sum_{t \in\{ t_{j_1},t_{j_2}\}}\mathbb{1}_{f_{t_i}(L,I) = f_{t}(L,I)} g_{t}(L,I), & \text{if } L = \emptyset,
\end{cases}
\end{equation}
where \( \mathbb{1}_{x=y} \) is the indicator function.

Let \( S \) be a trapping set in \( G_i \), and define \[S_0 = S \cap B_{t_i} \]  
\[ S_1= S\cap G_{t_{j_1}} \]
\[ S_2= S\cap G_{t_{j_2}} \]
We claim that $S$ is a $(L,I,t_i)$-type trapping set if and only if $S_1$ is a $(L,I',t_{j_1})$-type trapping set and $S_2$ is a $(L,I'',t_{j_2})$-type trapping set where $I = I' \oplus I'' \oplus \Gamma^o_{G[B_{t_i}]}(L)$. This claim follows from the fact that:
\begin{equation}
\label{a0_jn_1}
\begin{split}
    S \subset\mathcal{L}_i, S_1 \subset \mathcal{L}_{j_1},S_2 \subset \mathcal{L}_{j_2}
\end{split}
\end{equation}
\begin{equation}
\label{a0_jn_2}
\begin{split}
&S_1 \cap B_{t_{j_1}} =S\cap G_{t_{j_1}} \cap B_{t_{j_1}} \\&= 
     S \cap B_{t_i} =S_2 \cap B_{t_{j_2}}
\end{split}
\end{equation}
and
\begin{equation}   
\label{a0_jn_3}
\begin{split}
  &\Gamma_{G_i }^o(S)\\
  &=\Gamma_{G_i }^o(S_0\cup (S_1\setminus S_0)\cup (S_2\setminus S_0))\\
  &\overset{(b)}{=}\Gamma_{G_i }^o(S_0) \oplus \Gamma_{G_i }^o( S_1\setminus S_0) \oplus\Gamma_{G_i }^o( S_2\setminus S_0)\\
  &\overset{(c)}{=}\Gamma_{G_i }^o(S_0) \oplus \Gamma_{G_{j_1} }^o( S_1\setminus S_0) \oplus\Gamma_{G_{j_2} }^o( S_2\setminus S_0) \\&\overset{(d)}{=} \Gamma_{G_i }^o(S_0) \oplus \Gamma_{G_{j_1} }^o( S_1
  ) \oplus \Gamma_{G_{j_1} }^o(S_0) \oplus  \Gamma_{G_{j_2} }^o( S_2) \oplus \Gamma_{G_{j_2} }^o(S_0) 
\end{split}
 \end{equation}
The equality in (b) follows from the definitions of \(S_0\), \(S_1\), and \(S_2\). In particular, the sets \(S_0\), \(S_1 \setminus S_0\), and \(S_2 \setminus S_0\) are pairwise disjoint, and the result follows directly from Proposition~1. The equality in (c) is a consequence of the definition of a tree decomposition. Specifically, the set \(S_1 \setminus S_0 = S \cap (G_{t_{j_1}} \setminus B_{t_i})\) is adjacent only to check nodes in \(G_{t_{j_1}}\), which implies that \[\Gamma_{G_{i} }^o( S_1\setminus S_0)=\Gamma_{G_{j_1} }^o( S_1\setminus S_0)\]
The equality in \(d\) follows from the inclusio \(S_0 \subseteq S_1\), and again from Proposition~1.

According to the Proposition~2, we have
\begin{equation}
    \begin{split}
        \Gamma_{G_{i} }^o(S_0) = \Gamma_{G_{i}\setminus G_{j_1} }^o(S_0) \oplus \Gamma_{G_{i}\setminus G_{j_2} }^o(S_0) \oplus \Gamma_{G[B_{t_i}] }^o(S_0)
    \end{split}
\end{equation}
\begin{equation}
    \begin{split}
      \Gamma_{G_{j_1} }^o(S_0) &= \Gamma_{G_{j_1}\setminus G[B_{t_i}] }^o(S_0) \oplus \Gamma_{G[B_{t_i}] }^o(S_0) \\
      &=\Gamma_{G_{i}\setminus G_{j_2} }^o(S_0) \oplus \Gamma_{G[B_{t_i}] }^o(S_0)
    \end{split}
\end{equation}
\begin{equation}
    \begin{split}
      \Gamma_{G_{j_2} }^o(S_0) = \Gamma_{G_{i}\setminus G_{j_1} }^o(S_0) \oplus \Gamma_{G[B_{t_i}] }^o(S_0)
    \end{split}
\end{equation}
Substituting these equations into~(\ref{a0_jn_3}), we obtain
\begin{equation}
    \Gamma_{G_i }^o(S) =\Gamma_{G_{j_1}}^o(S_1) \oplus \Gamma_{G_{j_2}}^o(S_2) \oplus \Gamma^o_{G[B_{t_i}]}(S_0).
    \label{a0_jn_4}
\end{equation}

Taken together, (\ref{a0_jn_1}), (\ref{a0_jn_2}), and (\ref{a0_jn_4}) prove the claim.
\begin{itemize}
    \item If \(L \neq \emptyset\), then $S_1$ and \(S_2\) cannot be empty. Consequently, by the claim, we have$ f_{t_{i}}(L,I) =F_{t_{i}}(L,I), g_{t_{i}}(L,I), =G_{t_{i}}(L,I). $
    \item If \(L = \emptyset\), then $S_1$ and \(S_2\) may also be empty. In this situation, \(f_{t_{j_1}}(\emptyset, \emptyset)\) and $g_{t_{j_1}}(\emptyset, \emptyset)$ represents the size and the number of minimum \((a,0)\) trapping sets in \(G_{j_1}\), excluding the case of the empty set $S_1$. Therefore, this case requires special consideration, which leads to \(f_{t_i}(L,I)=\min \{F_{t_i}(L,I),f_{t_{j_1}}(L,I), f_{t_{j_2}}(L,I)\}\) and \(g_{t_i}(L,I)=\mathbb{1}_{f_{t_i}(L,I) = F_{t_i}(L,I)}G_{t_i}(L,I)+\sum_{t \in\{ t_{j_1},t_{j_2}\}}\mathbb{1}_{f_{t_i}(L,I) = f_{t}(L,I)} g_{t}(L,I)\).
\end{itemize}

For each pair \( (L \subset \mathcal{L}^B_{i}, R \subset \mathcal{R}^B_{i}) \), it takes \( O(2^{k}) \) time to obtain the minimum value in (\ref{a0_join_node_f1}). Therefore, the computational complexity for each join node is \( O(4^{k}) \), while for each other type of node, it is \( O(2^k) \). Given that a nice tree decomposition \( (T, (B_t)_{t \in T}) \) of the Tanner graph \( G \) contains \( O(n) \) nodes in \( T \), the overall time complexity of the algorithm is \( O(n 4^{k}) \).
\end{proof}

Based on Theorem \ref{main_a0}, if the pathwidth of a Tanner graph is known, we have the following corollary.

\begin{corollary}
Let \( G \) be a Tanner graph with pathwidth \( pw(G) = k\), corresponding to an LDPC code of length \( n \). Then, the size and the number of the smallest $(a,0)$ trapping set in $G$ can be determined in time \( O(n2^{k}) \). 
\end{corollary}
\begin{proof}
The proof proceeds similarly to that of Theorem \ref{main_a0}. However, in the case where \(T\) is a path, there exists no \(t \in T\) that is a join node. For all other types of nodes, the algorithm requires \(O(2^k)\) time to compute $f_t$ and $g_t$, which leads to an overall computational complexity of \( O(n2^{k}) \) for the algorithm.
\end{proof}

By executing the algorithm in Theorem \ref{main_a0} several times, we can identify an $(a,0)$ trapping set with minimum $a$ in \( G \). This leads to the following corollary:
\begin{corollary}
Let \( G \) be a Tanner graph with treewidth \( tw(G)=k \), corresponding to an LDPC code of length \( n \). We can identify an \((a,0)\) trapping sets with minimum $a$ in \( G \) in time \( O(n(a+1)4^{k}) \).
\end{corollary}
\begin{proof}
By repeatedly applying the algorithm from Theorem 1 as described in Algorithm 1, a minimum \((a, 0)\) trapping set can be identified. The MTS algorithm is a slight modification of the algorithm presented in Theorem 1. It operates on the graph \( G \) together with a tree decomposition \( (T,(B_t)_{t\in T}) \), and returns a node \( u \) such that, for some \( t_i \in T \) and its child \( t_j \), $B_{t_i} = B_{t_j} \setminus \{u\}$, and the corresponding value \( f_{t_i}(v,\emptyset) \) reaches \( a \) for the first time during the execution of the algorithm.
\end{proof}
\begin{algorithm}
\caption{FMTS($G,T,(B_t)_{t\in T}$)}
Let $a_0$ be the minimum size of an $(a,0)$ trapping set, as computed by the algorithm in Theorem 1\;
Add a variable node \( v \) to \( G \) such that \( v \) is not connected to any check node and appears in all \( B_t \), \( t \in T \)\;
$\mathcal{A}$ =[]\;
$a=a_0+1$\;
\ForEach{$i=1:a_0$}
{
    $u=MTS(G,T,(B_t)_{t\in T},v,a)$\;
    
        $\mathcal{A}$ = [$\mathcal{A}, u$]\; 
        $a = a-1$\;
        \If{$|\mathcal{A}|=a_0$}
        {
        break\;
        }
        Re-connect \( v \), where \( \Gamma_G^o(v) = \Gamma_G^o(u) \oplus \Gamma_G^o(v) \)\;
    Remove the node $u$ from $G$ and all $B_t, t\in T$\;
}
Return $\mathcal{A}$\;
\end{algorithm}

\section{Computer The Size and Number of the Smallest Trapping Set with \( b > 0 \) in Codes with Bounded Treewidth}
\label{section_ab}

As a generalization of Section \ref{section_a0}, this section presents an algorithm to find an \((a, b)\)-trapping set with the minimum \( a \) in Tanner graph \( G \), given \( b > 0 \), in linear complexity with respect to the code length. Furthermore, we also demonstrate that the number of smallest trapping sets with \( b > 0 \) contained in \( G \) can be computed in linear complexity with respect to the code length.

First, we provide the generalized form of the algorithm in Theorem \ref{main_a0} to compute the size and the number of the smallest trapping sets with \( b > 0 \) in \( G \), along with its proof.

\begin{theorem}
\label{main_ab}
   Let \( G \) be a Tanner graph with treewidth \( tw(G)=k \), corresponding to an LDPC code of length \( n \). Then, for a given $b> 0$, the size and the number of the smallest $(a,b)$ trapping set in $G$ can be determined in time \( O(n (b+1)^24^{k}) \). 
\end{theorem}
\begin{proof}
A nice tree decomposition of \( G \) is given by \( (T, (B_t)_{t \in T}) \), with root \( r \) and width \( k \). We adopt a dynamic programming algorithm, similar to that in Theorem~\ref{main_a0}, to solve the Minimum Trapping Set problem. The algorithm likewise processes the nodes of \(T\) in a bottom-up order \(t_1, t_2, \ldots, t_r\), such that every child of a node \(t_i\) appears earlier in the order. The objective is achieved by sequentially updating the function values \(f_{t_i}\) and \(g_{t_i}\) at each node $t_i \in T$. Compared with the algorithm presented in Theorem~1, the primary distinction lies in the redefinition of the functions \(f_{t_i}\) and \(g_{t_i}\).

\begin{definition} Let \(S\) be a trapping set in \(G\), \(L\) a set of variable nodes, \(R\) a set of check nodes, $d$ an integer. We say that \(S\) is an \((L, R, d, t_i)\)-type trapping set if the following conditions hold:
\begin{enumerate}
    \item $S \subset \mathcal{L}_i$;
\item \( S \cap B_{t_i} = L \);
\item $\Gamma^o_{G_i}(S) \cap B_{t_i} = R$,
\item $\Gamma^o_{G_i}(S) \setminus B_{t_i}| = d$.
\end{enumerate}
\end{definition}

For each node \( t_i \in T \), we define two functions as follows,
\[
f_{t_i} : (L ,R,d) \mapsto f_{t_i}(L, R,d) \in \mathbb{Z},
\]
and 
\[
g_{t_i}: (L ,R,d) \mapsto g_{t_i}(L, R,d) \in \mathbb{Z}.
\]
where $L \subset \mathcal{L}^B_i$ is a set of variable nodes in $B_{t_i}$, $R\subset \mathcal{R}^B_i$ is a set of check nodes in $B_{t_i}$. The value $f_{t_i}(L, R,d)$ denotes the minimum size of a $(L,R,d,t_i)$-type trapping set, while $g_{t_i}(L, R,d)$ counts the number of such minimum-size trapping sets. Similarly, after processing root node $t_r$, \(f_{t_r}(\emptyset,\emptyset,b)\) gives the minimum size of an \((a,b)\) trapping set contained in $G$, while \(g_{t_r}(\emptyset,\emptyset,b)\) gives the number of minimum \((a,b)\) trapping set contained in $G$. The detailed update rules and complexity analysis are provided in the sequel.

\noindent \textbf{Leaf node}: Suppose $t_i$ is a leaf node in $T$. Then,
\[
    f_{t_i}(L,R,d) = +\infty, 
\]
and
\[
    g_{t_i}(L,R,d) = 0, 
\]
for $L=\emptyset$, $R =\emptyset$, and $d=0$. It takes \( O(1) \) time to compute all \( f_{t_i}(L, R, d) \) and \( g_{t_i}(L, R, d) \) for each leaf node \( t_i \).

\noindent \textbf{introduced variable node}: Suppose \( t_i \) is an introduced node with a unique child \( t_j \), and that the corresponding bags satisfy \( B_{t_i} = B_{t_j} \cup \{v\} \), where \( v \) is a variable node. We can compute \( f_{t_i} \) and $g_{t_i}$ using the equations (\ref{ab_introduce_variable_f}) and (\ref{ab_introduce_veriable_g}) for all variable node set $L \subset \mathcal{L}^B_i$, check nodes set $R \subset \mathcal{R}^B_i$, and $d \in \{0,1,\ldots,b\}$.

\begin{figure*}
    \begin{equation}
\label{ab_introduce_variable_f}
   f_{t_i}(L,R,d) = \begin{cases}
        f_{t_j}(L,R,d),  &\text{if } v \notin L; \\ 
        1,  &\text{if } L = \{v\} , d =0, \text{ and } R = \Gamma_{G_i}^o(v); \\
        f_{t_j}(L\backslash  \{v\},R \oplus \Gamma_{G_i}^o(v),d) + 1,  &\text{otherwise.}
   \end{cases}
\end{equation}
\end{figure*}
\begin{figure*}
\begin{equation}
\label{ab_introduce_veriable_g}
   g_{t_i}(L,R,d) = \begin{cases}
        g_{t_j}(L,R,d),  &\text{if } v \notin L; \\ 
        1,  &\text{if } L = \{v\} , d =0, \text{ and } R = \Gamma_{G_i}^o(v); \\
        g_{t_j}(L\backslash  \{v\},R \oplus \Gamma_{G_i}^o(v),d),  &\text{otherwise.}
   \end{cases}
\end{equation}
\end{figure*}
Next, we prove the correctness of (\ref{ab_introduce_variable_f}) and (\ref{ab_introduce_veriable_g}). Let \( S \) be a trapping set in \( G \). For each \( (L, R , d )\), the following conclusion holds.
\begin{itemize}
    \item For the case where $v \notin L$, according to \eqref{a0_ivn_1} $\sim$ \eqref{a0_ivn_3} when $v \notin S $, we have 
\begin{equation}
\label{ab_ivn_1}
\begin{split}
    S \subset \mathcal{L}_j\Leftrightarrow S \subset  \mathcal{L}_i,
\end{split}
\end{equation}
\begin{equation}
\label{ab_ivn_2}
\begin{split}
    S \cap B_{t_j} =S\cap B_{t_i},
\end{split}
\end{equation}
\begin{equation}   
\label{ab_ivn_3}
\Gamma_{G_{j}}^o(S) \cap B_{t_j} = \Gamma_{G_i}^o(S) \cap B_{t_i},
 \end{equation}
 and
 \begin{equation}   
\label{ab_ivn_4}
\Gamma_{G_{j}}^o(S) \setminus B_{t_i} = \Gamma_{G_j}^o(S) \setminus B_{t_i}.
 \end{equation}

When \( S \) is a \( (L, R, d,t_i) \)-type trapping set, we have \( S \cap B_{t_i} = L \), and \( v \notin L \), thus \( v \notin S \). When \( S \) is a \( (L, R, d,t_j) \)-type trapping set, we have \( S \subset\mathcal{L}_j\), thus \( v \notin S \). Therefore, $S$ is a \( (L, R,d,t_i) \)-type trapping set if and only if $S$ is also a \( (L, R,d,t_j) \)-type trapping set, which leads to
$f_{t_i}(L, R,d) = f_{t_j}(L, R,d) $ and $g_{t_i}(L, R,d) = g_{t_j}(L, R,d)$. 

 \item For the case where \( L = \{v\} ,d=0\) and \( R = \Gamma^o_{G_i}(v) \), \( \{v\} \) is a \( (L, R,0,t_i) \)-type trapping set. In this case, we have \( f_{t_i}(L, R,0) = 1 \) and \( g_{t_i}(L, R,0) = 1 \).

\item For other cases, \( S \) has the form \( S = \{v\} \cup S' \), where $v \notin S'$. We claim $S$ is a \( (L, R,d,t_i) \)-type trapping set if and only if $S'$ is a \( (L\setminus \{v\}, R \oplus \Gamma^o_{G_i}(v),d,t_j) \)-type trapping set, which leads to
$f_{t_i}(L,R,d) = f_{t_j}(L\setminus \{v\} , R \oplus \Gamma^o_{G_i}(v),d)+1.$ and $g_{t_i}(L,R,d) = g_{t_j}(L\setminus \{v\} , R \oplus \Gamma^o_{G_i}(v),d)$. According to \eqref{a0_ivn_4} $\sim$ \eqref{a0_ivn_6}, we have 
\begin{equation}
\label{ab_ivn_5}
\begin{split}
    S' \subset\mathcal{L}_j\Leftrightarrow S \subset \mathcal{L}_i
\end{split}
\end{equation}
\begin{equation}
\label{ab_ivn_6}
\begin{split}
    S'\cap B_{t_j} =(S\cap B_{t_i}) \setminus \{v\}
\end{split}
 \end{equation}
\begin{equation}
\label{ab_ivn_7}
\begin{split}
    \Gamma^o_{G_j}(S')\cap B_{t_j} &= (\Gamma^o_{G_i}(S) \oplus \Gamma^o_{G_i}(v)) \cap B_{t_i}\\
    &=(\Gamma^o_{G_i}(S) \cap B_{t_i})\oplus (\Gamma^o_{G_i}(v) \cap B_{t_i})\\
    &\overset{(e)}{=}(\Gamma^o_{G_i}(S) \cap B_{t_i})\oplus \Gamma^o_{G_i}(v)
\end{split}
\end{equation}
and
\begin{equation}
\label{ab_ivn_8}
\begin{split}
    \Gamma^o_{G_j}(S')\setminus B_{t_j} &= (\Gamma^o_{G_i}(S) \oplus \Gamma^o_{G_i}(v)) \setminus B_{t_i}\\
    &=\Big(\Gamma^o_{G_i}(S) \setminus B_{t_i} \Big) \oplus \Big(\Gamma^o_{G_i}(v) \setminus B_{t_i}\Big)\\
    &\overset{(f)}{=}\Gamma^o_{G_i}(S) \setminus B_{t_i}
\end{split}
\end{equation}
The validity of $(e)$ and $(f)$ follows from the definition of treewidth: since \(v\) appears for the first time in \(B_{t_i}\), it can be adjacent only to check nodes within \(B_{t_i}\), implying that \(\Gamma^o_{G_i}(v) \subseteq B_{t_i}\). 

Clearly, based on equations (\ref{ab_ivn_5})$\sim$(\ref{ab_ivn_8}), the sufficiency of the claim is established. Conversely, to establish necessity, suppose \( S' \) is a \( (L \setminus {v}, R \oplus \Gamma^o_{G_i}(v),d, t_j) \)-type trapping set, then \( S \cap B_{t_i} = L \) or \( S \cap B_{t_i} = L\setminus\{v\} \). Since \( v \in S\), it follows that \( S \cap B_{t_i} = L \). Therefore, \( S \) is a \( (L, R, d,t_i) \)-type trapping set.  
\end{itemize}
It takes \( O((b+1)2^k) \) time for each introduced variable node to compute \( f_{t_i}(L, R,d) \) and \( g_{t_i}(L, R,d) \) for all \( (L, R,d) \).

\noindent \textbf{Forget variable node}: Suppose \( t_i \) is a forget node with a unique child \( t_j \), and that the corresponding bags satisfy \( B_{t_i} = B_{t_j} \setminus \{v\} \), where \( v \) is a variable node. We can compute \( f_{t_i} \) and \( g_{t_i} \) using the following equations for all $L \subset \mathcal{L}^B_i$, $R \subset \mathcal{R}^B_i$, and $d \in \{0,1,\ldots,b\}$:
\begin{equation}
\label{ab_forget_variable_f}
    f_{t_i}(L,R,d) = \min \{f_{t_j}(L,R,d),f_{t_j}(L\cup \{v\},R,d)\}.
\end{equation}
and
\begin{equation}
\label{ab_forget_variable_g}
    g_{t_i}(L,R,d) =  \sum_{\substack{P\in\{(L,R,d),(L\cup \{v\},R,d)\}\\f_{t_i}(L,R,d)=f_{t_j}(P) }} g_{t_j}(P).
\end{equation}

Let \( S \) be a trapping set in \( G \). For each \( (L,R,d)\), we claim that \( S \) is a \( (L, R,d,t_i) \)-type trapping set if and only if $S$ is either a \( (L, R,d,t_j) \)-type trapping set or a \( (L\cup \{v\}, R,d,t_j) \)-type trapping set, which leads to (\ref{ab_forget_variable_f}) and (\ref{ab_forget_variable_g}). According to \eqref{a0_fvn_1} $\sim$ \eqref{a0_fvn_3}, we have 
\begin{equation}
\label{ab_fvn_1}
  \begin{split}
    S \subset\mathcal{L}_j\Leftrightarrow S \subset \mathcal{L}_i
\end{split}  
\end{equation}
\begin{equation}
\label{ab_fvn_2}
  \begin{split}
    S \cap B_{t_j} &=(S\cap B_{t_i}) \cup (S \cap \{v\}),
\end{split}  
\end{equation}
\begin{equation}
\label{ab_fvn_3}
 \begin{split}
     \Gamma^o_{G_j}(S)\cap B_{t_i} =\Gamma^o_{G_i}(S)\cap B_{t_j}.
\end{split}   
\end{equation}
and
\begin{equation}
\label{ab_fvn_4}
 \begin{split}
     \Gamma^o_{G_j}(S)\setminus B_{t_i} =\Gamma^o_{G_i}(S)\setminus B_{t_j}.
\end{split}   
\end{equation}
Clearly, based on equations (\ref{ab_fvn_1})$\sim$(\ref{ab_fvn_4}), the sufficiency of the claim is established. Conversely, to establish necessity, suppose \( S \) is a \( (L, R, d,t_j) \)-type trapping set. Since \( v \notin L,v \in B_{t_j} \), and $S \cap B_{t_j} = L$, we have \( S \cap \{v\} = \emptyset \), which implies that \( S \cap B_{t_i} = L \). Therefore, \( S \) is a \( (L, R, d,t_i) \)-type trapping set. Similarly, suppose \( S \) is a \( (L \cup \{v\}, R, d,t_j) \)-type trapping set. According to \eqref{ab_fvn_2}, \( S \cap B_{t_i} = L \) or \( S \cap B_{t_i} = L \cup \{v\} \). Since \( v \notin B_{t_i} \), it follows that \( S \cap B_{t_i} = L \), and thus \( S \) is a \( (L, R,d, t_i) \)-type trapping set.

It takes \(  O((b+1)2^k) \) time for each forget variable node to compute \( f_{t_i}(L, R,d) \) and \( g_{t_i}(L, R,d) \) for all \( (L, R,d) \).

\noindent \textbf{introduced check node}: Suppose \( t_i \) is an introduced node with a unique child \( t_j \), and that the corresponding bags satisfy \( B_{t_i} = B_{t_j} \cup \{c\} \), where \( c \) is a check node. We can compute \( f_{t_i} \) and \( g_{t_i} \) using the following equations for all $L \subset \mathcal{L}^B_i$, $R \subset \mathcal{R}^B_i$, and $d \in \{0,1,\ldots,b\}$:

\begin{equation}
\label{ab_introduce_check_f}
f_{t_i}(L,R,d) = \begin{cases}
f_{t_j}(L,R\setminus\{c\},d), & \text{if } c \in R \text{, $E(c,L)$ is odd;}\\
f_{t_j}(L,R,d),& \text{if } c \notin R \text{, $E(c,L)$ is even;}\\
+\infty, & \text{otherwise},
\end{cases}
\end{equation}
\begin{equation}
\label{ab_introduce_check_g}
g_{t_i}(L,R,d) = \begin{cases}
g_{t_j}(L,R\setminus\{c\},d), & \text{if } c \in R \text{, $E(c,L)$ is odd;}\\
g_{t_j}(L,R,d),& \text{if } c \notin R \text{, $E(c,L)$ is even;}\\
0, & \text{otherwise},
\end{cases}
\end{equation}
where \( E(c,L) \) denotes the number of edges between \( c \) and \( L \) in \( G_i\).

Next, we prove the correctness of (\ref{ab_introduce_check_f}) and (\ref{ab_introduce_check_g}). Let \( S \) be a trapping set in \( G \). For each \( (L \subset \mathcal{R}^B_{i}, R \subset \mathcal{L}^B_{i},d \in \{0,1,\ldots,b\}) \), according to \eqref{a0_icn_1} $\sim$ \eqref{a0_icn_4}, we have
\begin{equation}
\label{ab_icn_1}
\begin{split}
   S \subset\mathcal{L}_j\Leftrightarrow S \subset \mathcal{L}_i
\end{split}
\end{equation}
\begin{equation}
\label{ab_icn_2}
\begin{split}
    S \cap B_{t_j} =  S \cap B_{t_i}
\end{split}
\end{equation}
\begin{equation}   
\label{ab_icn_3}
\begin{split}
 \Gamma_{G_{j}}^o(S)\cap B_{t_j} &=(\Gamma_{G_i}^o(S) \oplus  \Gamma_{G[c\cup S]}^o(S\cap B_{t_i}))\cap (B_{t_i}\setminus \{c\})\\
 &=\left((\Gamma_{G_i}^o(S) \cap B_{t_i})\setminus(\Gamma_{G_i}^o(S) \cap \{c\})\right)\\
 &\oplus \left(\Gamma_{G[c\cup S]}^o(S\cap B_{t_i}) \cap (B_{t_i}\setminus \{c\})\right)\\
 &=(\Gamma_{G_i}^o(S) \cap B_{t_i})\setminus(\Gamma_{G_i}^o(S) \cap \{c\})
\end{split}
 \end{equation}
 and
 \begin{equation}   
\label{ab_icn_4}
\begin{split}
 &\Gamma_{G_{j}}^o(S)\setminus B_{t_j}  =(\Gamma_{G_i}^o(S) \oplus  \Gamma_{G[c\cup S]}^o(S\cap B_{t_i}))\setminus(B_{t_i}\setminus \{c\})\\
 &=\Big((\Gamma_{G_i}^o(S) \oplus  \Gamma_{G[c\cup S]}^o(S\cap B_{t_i}))\setminus B_{t_i}\Big)\\
 &\cup \Big((\Gamma_{G_i}^o(S) \oplus  \Gamma_{G[c\cup S]}^o(S\cap B_{t_i})) \cap \{c\}\Big)\\
 &\overset{(g)}{=} \Big(\Gamma_{G_i}^o(S) \setminus B_{t_i}\Big) \oplus \Big(\Gamma_{G[c\cup S]}^o(S\cap B_{t_i}) \setminus B_{t_i}\Big)\\
 &=\Gamma_{G_i}^o(S) \setminus B_{t_i}.
\end{split}
 \end{equation}
The validity of $(g)$ follows from the definition of treewidth: $c$ can be adjacent only to variable nodes within \(B_{t_i}\), implying that \(E(c,S)=E(c, S\cap B_{t_i})\). Therefore, $c \notin \Gamma_{G_i}^o(S) \oplus  \Gamma_{G[c\cup S]}^o(S\cap B_{t_i})$, and thus $(\Gamma_{G_i}^o(S) \oplus  \Gamma_{G[c\cup S]}^o(S\cap B_{t_i})) \cap \{v\} =\emptyset$.
 
 \begin{itemize}
\item If $E(c,L)$ is odd and $c \in R$, according to (\ref{ab_icn_1})–(\ref{ab_icn_4}), we have that $S$ is a \( (L, R,d,t_i) \)-type trapping set if and only if $S$ is a \( (L, R\setminus\{c\},d,t_j) \)-type trapping set. Consequently, $f_{t_i}(L, R,d) = f_{t_j}(L, R\setminus\{c\},d) $ and $g_{t_i}(L, R,d) = g_{t_j}(L, R\setminus\{c\},d)$. 
 
\item If $E(c,L)$ is even and $c \notin R$, according to (\ref{ab_icn_1})–(\ref{ab_icn_4}), we have that $S$ is a \( (L, R,d,t_i) \)-type trapping set if and only if $S$ is a \( (L, R,d,t_j) \)-type trapping set. Consequently, $f_{t_i}(L, R,d) = f_{t_j}(L, R,d) $ and $g_{t_i}(L, R,d) = g_{t_j}(L, R,d)$.

\item If $E(c,L)$ is odd and $c \notin R$, we claim that there does not exist an $(L,R,d,t_i)$-type trapping set $S$ in $G_i$. Otherwise, since \(E(c,L=S\cap B_{t_i})\) is odd, we have $E(c,S)$ is odd and \(
c \in \Gamma^{o}_{G_i}(S).
\)
Therefore, \(c \in \Gamma^{o}_{G_i}(S) \cap B_{t_i}\), which contradicts the assumption that \(c \notin R\). 

\item If $E(c,L)$ is even and $c \in R$, we claim that there does not exist an $(L,R,d,t_i)$-type trapping set $S$ in $G_i$. Otherwise, since \(E(c,L=S\cap B_{t_i})\) is even, we have $E(c,S)$ is even and \(
c \notin \Gamma^{o}_{G_i}(S).
\)
Therefore, \(c \notin \Gamma^{o}_{G_i}(S) \cap B_{t_i}\), which contradicts the assumption that \(c \in R\).

 \end{itemize}

It takes \( O((b+1)2^k) \) time for each introduced check node to compute \( f_{t_i}(L, R,d) \) and \( g_{t_i}(L, R,d) \) for all \( (L, R,d) \).

\noindent \textbf{Forget check node}: Suppose \( t_i \) is a forget node with a unique child \( t_j \), and that the corresponding bags satisfy \( B_{t_i} = B_{t_j} \setminus \{c\} \), where \( c \) is a check node. We can compute \( f_{t_i} \) and \( g_{t_i} \) using the following equations for all $L \subset \mathcal{L}^B_i$, $R \subset \mathcal{R}^B_i$, and $d \in \{0,1,\ldots,b\}$:

\begin{equation}
\label{ab_forget_check_f}
    f_{t_i}(L,R,d) = \min\{f_{t_j}(L,R,d),f_{t_j}(L,R\cup \{c\},d-1)\},
\end{equation}
\begin{equation}
\label{ab_forget_check_g}
    g_{t_i}(L,R,d) =  \sum_{\substack{P\in\{(L,R,d),(L,R\cup \{c\},d-1)\}\\f_{t_i}(L,R,d)=f_{t_{j}}(P) }} g_{t_j}(P).
\end{equation}

Let \( S \) be a trapping set in \( G \). For each pair \( (L \subset \mathcal{R}^B_{i}, R \subset \mathcal{L}^B_{i},d \in \{0,1,\ldots,b\}) \), we claim that $S$ is a \( (L, R,d,t_i) \)-type trapping set if and only if $S$ is a \( (L, R,d,t_j) \)-type trapping set or $(L,R\cup \{c\},d-1,t_j)$-type trapping set, which leads to \eqref{ab_forget_check_f} and \eqref{ab_forget_check_g}. According to \eqref{a0_fcn_1} $\sim$ \eqref{a0_fcn_3}, we have 
\begin{equation}
\label{ab_fcn_1}
\begin{split}
   S \subset\mathcal{L}_j\Leftrightarrow S \subset \mathcal{L}_i
\end{split}
\end{equation}
\begin{equation}
\label{ab_fcn_2}
\begin{split}
    S \cap B_{t_j} =  S \cap B_{t_i}
\end{split}
\end{equation}
\begin{equation}   
\label{ab_fcn_3}
\begin{split}
 \Gamma_{G_{j}}^o(S) \cap B_{t_j} &= \Gamma_{G_i }^o(S) \cap (B_{t_i}\cup \{c\})\\
 &=(\Gamma_{G_{i}}^o(S)\cap B_{t_i}) \cup (\Gamma_{G_{i}}^o(S) \cap \{c\})\
\end{split}
 \end{equation}
\begin{equation}   
\label{ab_fcn_4}
\begin{split}
 \Gamma_{G_{j}}^o(S) \setminus B_{t_j} &= \Gamma_{G_i }^o(S) \setminus (B_{t_i}\cup \{c\})\\
 &=(\Gamma_{G_{i}}^o(S)\setminus B_{t_i}) \setminus\{c\}
\end{split}
 \end{equation}

Clearly, based on equations (\ref{ab_fcn_1})$\sim$(\ref{ab_fcn_4}), the sufficiency of the claim is established. Conversely, to establish necessity, suppose \( S \) is a \( (L, R, d,t_j) \)-type trapping set, then
$c \notin R = \Gamma_{G_{j}}^o(S) \cap B_{t_j}$. According to \eqref{ab_fcn_3}, we have $c \notin \Gamma_{G_{i}}^o(S)$, $\Gamma_{G_{i}}^o(S)\cap B_{t_i} =R$, and $|\Gamma_{G_{i}}^o(S)\setminus B_{t_i}|=b$. Therefore, \( S \) is a \( (L, R, d,t_i) \)-type trapping set. Similarly, suppose \( S \) is a \( (L, R\cup \{c\}, d-1,t_j) \)-type trapping set. According to \eqref{ab_fcn_3}, \( \Gamma_{G_{i}}^o(S)\cap B_{t_i} = R \) or \( \Gamma_{G_{i}}^o(S)\cap B_{t_i} = R \cup \{c\} \). Since \( c \notin B_{t_i} \), it follows that \( \Gamma_{G_{i}}^o(S)\cap B_{t_i} = R \) and $c \in \Gamma_{G_{i}}^o(S)$. Therefore, $|(\Gamma_{G_{i}}^o(S)\setminus B_{t_i})|=b$ and \( S \) is a \( (L, R,d, t_i) \)-type trapping set.

It takes \( O((b+1)2^k) \) time for each forget check node to compute \( f_{t_i}(L, R, d) \) and \( g_{t_i}(L, R, d) \) for all \( (L, R, d) \).

\noindent \textbf{Join node}: Suppose \( t_i \) is a join node, and \( t_{j_1} \) and \( t_{j_2} \) be its two child nodes with $B_{t_i} = B_{t_{j_1}} = B_{t_{j_2}}$. We define the following two functions for all $L \subset \mathcal{L}^B_i$, $R \subset \mathcal{R}^B_i$, and $d \in \{0,1,\ldots,b\}$:
\begin{equation}
\label{ab_join_node_f1}
F_{t_i}(L,R,d) = \min_{\substack{R',R''\\d',d''}}\{f_{t_{j_1}}(L,R',d')+ f_{t_{j_2}}(L,R'',d'') - |L|\}.
\end{equation}
\begin{equation}
\label{ab_join_node_f2}
\begin{split}
  &G_{t_i}(L,R,d) = \sum_{\substack{R',R'',d',d''}} g_{t_{j_1}}(L,R',d') \times g_{t_{j_2}}(L,R'' ,d'').
\end{split}
\end{equation}
where (\ref{ab_join_node_f1}) takes the minimum over all \( R', R'',d',d'' \) satisfying \( R = R' \oplus R'' \oplus \Gamma^o_{G[B_t]}(L)\) and $d = d'+d''$, (\ref{ab_join_node_f2}) takes the sum over all \( R',R'',d',d'' \) that attain the minimum value in (\ref{ab_join_node_f1}). 
\begin{figure*}
\begin{equation}
\label{a0_join_node_f2}
f_{t_i}(L,R,d) =
\begin{cases}
     F_{t_{i}}(L,R,d), \text{\qquad\qquad\qquad\qquad\qquad\;\;\;\; if } L \neq \emptyset \\
     \min \{F_{t_i}(L,R,d),f_{t_{j_1}}(L,R,d), f_{t_{j_2}}(L,R,d) \}, \text{if } L = \emptyset
\end{cases}
\end{equation}
\end{figure*}
\begin{figure*}
\begin{equation}
\label{a0_join_node_g2}
g_{t_i}(L,R,d) =
\begin{cases}
     G_{t_i}(L,R,d),& \text{if } L \neq \emptyset; \\
     \mathbb{1}_{f_{t_i}(L,R,d) = F_{t_i}(L,R,d)}G_{t_i}(L,I)
     \\+\sum_{t \in\{ t_{j_1},t_{j_2}\}}\mathbb{1}_{f_{t_i}(L,R,d) = f_{t}(L,R,d)} g_{t}(L,R,d), & \text{if } L = \emptyset,
\end{cases}
\end{equation}
\end{figure*}

Let \( S \) be a trapping set in \( G_i \), and define \[S_0 = S \cap B_{t_i} \]  
\[ S_1= S\cap G_{t_{j_1}} \]
\[ S_2= S\cap G_{t_{j_2}} \]
We claim that $S$ is a $(L,R,d,t_i)$-type trapping set if and only if $S_1$ is a $(L,R',d',t_{j_1})$-type trapping set and $S_2$ is a $(L,R'',d'',t_{j_2})$-type trapping set where $R = R' \oplus R'' \oplus \Gamma^o_{G[B_{t_i}]}(L)$. According to \eqref{a0_jn_1} , \eqref{a0_jn_2}, and \eqref{a0_jn_4}, we have 

\begin{equation}
\label{ab_jn_1}
\begin{split}
    S \subset\mathcal{L}_i, S_1 \subset \mathcal{L}_{j_1},S_2 \subset \mathcal{L}_{j_2}
\end{split}
\end{equation}
\begin{equation}
\label{ab_jn_2}
\begin{split}
&S \cap B_{t_i}=S_1 \cap B_{t_{j_1}} = S_2 \cap B_{t_{j_2}}
\end{split}
\end{equation}
\begin{equation}   
\label{ab_jn_3}
\begin{split}
  &\Gamma_{G_i }^o(S)\cap B_{t_i} \\
  &=\left(\Gamma_{G_{j_1}}^o(S_1) \oplus \Gamma_{G_{j_2}}^o(S_2) \oplus \Gamma^o_{G[B_{t_i}]}(S_0)\right)\cap B_{t_i}\\
  &=(\Gamma_{G_{j_1}}^o(S_1) \cap  B_{t_{j_1}}) \oplus (\Gamma_{G_{j_2}}^o(S_2) \cap  B_{t_{j_2}}) \oplus (\Gamma^o_{G[B_{t_i}]}(S_0)\cap  B_{t_i})
\end{split}
 \end{equation}
 and
 \begin{equation}   
\label{ab_jn_4}
\begin{split}
  &\Gamma_{G_i }^o(S)\setminus B_{t_i} \\
  &=\left(\Gamma_{G_{j_1}}^o(S_1) \oplus \Gamma_{G_{j_2}}^o(S_2) \oplus \Gamma^o_{G[B_{t_i}]}(S_0)\right)\setminus B_{t_i}\\
  &=(\Gamma_{G_{j_1}}^o(S_1) \setminus  B_{t_i}) \oplus (\Gamma_{G_{j_2}}^o(S_2) \setminus  B_{t_i}) \oplus (\Gamma^o_{G[B_{t_i}]}(S_0)\setminus  B_{t_i}) \\
  &= (\Gamma_{G_{j_1}}^o(S_1) \setminus  B_{t_{j_1}}) \oplus (\Gamma_{G_{j_2}}^o(S_2) \setminus  B_{t_{j_2}})
\end{split}
 \end{equation}
 According to the definition of tree decomposition, we have: \(
(G_{j_1} \setminus B_{t_{j_1}}) \cap (G_{j_2} \setminus B_{t_{j_2}}) = \emptyset.
\) This leads to the conclusion that \( |\Gamma_{G_i }^o(S)\setminus B_{t_i}|=|(\Gamma_{G_{j_1}}^o(S_1) \setminus  B_{t_{j_1}})|+|(\Gamma_{G_{j_2}}^o(S_2) \setminus  B_{t_{j_2}})|\). Together with (\ref{ab_jn_1}) $\sim$ (\ref{ab_jn_3}), this proves the claim.
\begin{itemize}
    \item If \(L \neq \emptyset\), then $S_1$ and \(S_2\) cannot be empty. Consequently, by the claim, we have$ f_{t_{i}}(L,R,d) =F_{t_{i}}(L,R,d), g_{t_{i}}(L,R,d), =G_{t_{i}}(L,R,d). $
    \item If \(L = \emptyset\), then $S_1$ and \(S_2\) may also be empty. In this situation, \(f_{t_{j_1}}(\emptyset, \emptyset,0)\) and $g_{t_{j_1}}(\emptyset, \emptyset,0)$ represents the size and the number of minimum \((a,0)\) trapping sets in \(G_{j_1}\), excluding the case of the empty set $S_1$. Therefore, this case requires special consideration, which leads to \(f_{t_i}(L,R,d)=\min \{F_{t_i}(L,R,d),f_{t_{j_1}}(L,R,d), f_{t_{j_2}}(L,R,d)\}\) and \(g_{t_i}(L,R,d)=\mathbb{1}_{f_{t_i}(L,R,d) = F_{t_i}(L,R,d)}G_{t_i}(L,R,d)+\sum_{t \in\{ t_{j_1},t_{j_2}\}}\mathbb{1}_{f_{t_i}(L,R,d) = f_{t}(L,R,d)} g_{t}(L,R,d)\).
\end{itemize}

For each \( (L \subset \mathcal{L}^B_{i}, R \subset \mathcal{R}^B_{i},d \in \{0,1,\ldots,b\}) \), it takes \( O(b2^{k}) \) time to obtain the minimum value in (\ref{ab_join_node_f1}). Therefore, the computational complexity for each join node is \( O(b^22^{2k}) \), while for each other type of node, it is \( O(b2^k) \). Given that a nice tree decomposition \( (T, (B_t)_{t \in T}) \) of the Tanner graph \( G \) contains \( O(n) \) nodes in \( T \), the overall time complexity of the algorithm is \( O(n b^22^{2k}) \).
\end{proof}

Similar to Section \ref{section_a0}, we apply the algorithm from Theorem \ref{main_ab} to find an \((a,b)\)-trapping set with the minimum \(a\) in linear time complexity. Furthermore, we can derive the computational complexity of determining the size and the number of the smallest trapping sets with a given $b>0$ under bounded pathwidth constraints.

\begin{corollary}
Let \( G \) be a Tanner graph with treewidth \( tw(G)=k \), corresponding to an LDPC code of length \( n \). We can identify an \((a,b)\) trapping sets with minimum $a$ in \( G \) in time \( O(n(a+1)(b+1)^3 4^{k}) \).
\end{corollary}
\begin{proof}
By repeatedly applying the algorithm from Theorem \ref{main_ab} as described in Algorithm 2, a minimum \((a, b)\) trapping set can be identified. The GMTS algorithm is a slight modification of the algorithm presented in Theorem \ref{main_ab}. It operates on the graph \( G \) together with a tree decomposition \( (T,(B_t)_{t\in T}) \), and returns a node \( u \) such that, for some \( t_i \in T \) and its child \( t_j \), $B_{t_i} = B_{t_j} \setminus \{u\}$, and the corresponding value \( f_{t_i}(v,\mathcal{B},b) \) reaches \( a \) for the first time during the execution of the algorithm.
\end{proof}
\begin{algorithm}
\caption{GFMTS($G,T,(B_t)_{t\in T},b$)}
Let $a_0$ be the minimum size of an $(a,b)$ trapping set, as computed by the algorithm in Theorem 1\;
Add a variable node \( v \) to \( G \) such that \( v \) is not connected to any check node and appears in all \( B_t \), \( t \in T \)\;
$\mathcal{A}$ =[], $\mathcal{B}$ =[]\;
$a=a_0+1$\;
\ForEach{$i=1:a_0+b$}
{

    $u=GMTS(G,T,(B_t)_{t\in T},v,\mathcal{B},b,a)$\;
    \uIf{$u$ is a check node}
    {

        $\mathcal{B}$ =[$\mathcal{B}, u$]\;
        $b=b-1$\;
    }
    \Else
    {
        $\mathcal{A}$ = [$\mathcal{A}, u$]\; 
        $a = a-1$\;
        \If{$|\mathcal{A}|=a_0$}
        {
        break\;
        }
        $\mathcal{B}'=\Gamma_G^o(u)\cap \mathcal{B}$\;
        $\mathcal{B} = \mathcal{B} \setminus \mathcal{B}'$\;
        Re-connect \( v \), where \( \Gamma_G^o(v) = (\Gamma_G^o(u) \setminus \mathcal{B}') \oplus \Gamma_G^o(v) \)\;
    }
    Remove the node $u$ from $G$ and all $B_t, t\in T$\;
}
Return $\mathcal{A}$\;
\end{algorithm}

\begin{corollary}
Let \( G \) be a Tanner graph with pathwidth \( pw(G) = k\), corresponding to an LDPC code of length \( n \). Then, for a given $b$, the size and the number of the smallest $(a,b)$ trapping set in $G$ can be determined in time \( O(n (b+1)2^{k}) \). 
\end{corollary}

\section{Simulation Results}
\label{section_algorithm}
In this section, we successfully apply the algorithm to spatially coupled LDPC codes, for which a tree decomposition, or even a path decomposition, can be easily found. To verify the correctness of the algorithm, we apply it to short spatially coupled LDPC codes, where all small trapping sets can be enumerated using an exhaustive search algorithm.

\begin{figure}[htbp]
\centering
\includegraphics[width=0.3\textwidth]{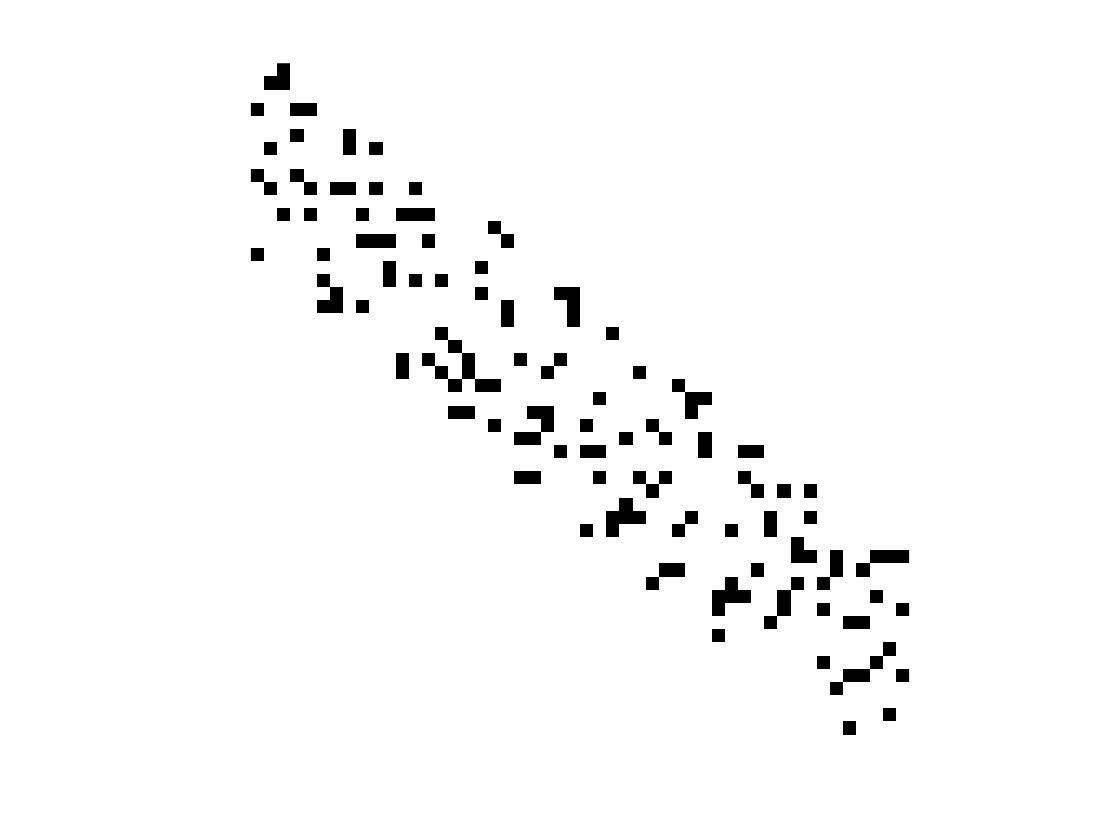}
\caption{A randomly generated spatially coupled LDPC code. The variable nodes have a regular degree of 3. The black regions indicate positions in the parity-check matrix where the entries are 1, while other positions are 0.}
\label{H}
\end{figure}

Figure \ref{H} presents a randomly generated spatially coupled LDPC code $G$, where the uncoupled subcode corresponds to a \(4 \times 5\) matrix. The coupling length is 10, and the coupling width (as defined in \cite{mitchell2015spatially}) is 3. We can easily obtain a path decomposition with a width equal to the product of the number of check nodes in the uncoupled subcode and the coupling width plus one. This is achieved by placing each variable node along with all its neighboring check nodes into the same bag. Furthermore, a nice rooted tree decomposition can be achieved by adding an additional bag to the structure. 

By applying the algorithm in Theorem 1 to the graph \( G \), we determine that the size of the smallest \((a,0)\) trapping set is 10, and there is one \((10,0)\) trapping set in \( G \). Algorithm 1 identifies a \((10,0)\) trapping set \((v_{12},v_{14},v_{20},v_{21},v_{22},v_{24},v_{25},v_{27},v_{30},v_{34})\), which is consistent with the result obtained via exhaustive search, thereby validating the correctness and effectiveness of the proposed algorithm. Furthermore, on a platform with an AMD Ryzen7 4800H CPU, the algorithm in Theorem 1 executed in 0.57 hours, while the exhaustive search algorithm took 50.13 hours, demonstrating an approximately 88-fold improvement in computational efficiency. As the code length and the minimum value of \( a \) increase, our algorithm will continue to outperform the exhaustive search, with a time complexity of \( O(n) \) compared to the \( O(n^a) \) complexity of the exhaustive search.



\section{Conclusion}
\label{section_conclusion}

For any fixed \( b \ge 0 \), we have shown that for LDPC codes with bounded treewidth, the minimum size of an \( (a,b) \) trapping set, the number of minimum trapping sets, and a corresponding trapping set itself can all be computed in time linear in the code length.

\bibliographystyle{ieeetr}
\bibliography{reference}

\end{document}